\documentclass[preprint2]{aastex}

\shorttitle{A view of the M81 galaxy group via $H\alpha$ window}
\shortauthors{Karachentsev \& Kaisin}
\begin{document}

\title{A view of the M81 galaxy group via the $H\alpha$ window}
\author{Igor D.\ Karachentsev
\and Serafim S.\ Kaisin}
\affil{Special Astrophysical Observatory, Russian Academy
	  of Sciences, N.\ Arkhyz, KChR, 369167, Russia}
\email{ikar@luna.sao.ru}
\begin{abstract}
We present images for 36 galaxies of the M81 group obtained in the
$H\alpha$ line. Estimates of the $H\alpha$ flux and star formation rate,
SFR, are avialable now for all the known members of the group with
absolute magnitudes down to  $M_B = -10^m$.

  The character of distribution of the galaxies over three paremeters:
$M_B$, SFR, and total hydrogen mass permits us to draw the following
conclusions as to evolution status of the group population. a) Spiral
and irregular type galaxies would have time to generate their luminosity
(baryon mass) during the cosmological time $T_0=13.7$ Gyr, but dwarf
spheroidal objects are capable of reproducing only $\sim$5\% of their
observed luminosity. b) S and Im,BCD galaxies possess the supply of gas
sufficient to maintain their observed SFRs during only next (1/4 -
 1/3)$T_0$ years, while dIr and dSph populations have the mean gas
depletion time about 3 $T_0$. c) There is indirect evidence that the star
formation in Im, BCD and dIr galaxies proceeds in a mode of vigorous burst
activity rather than in the form of a sluggish process. We note the
dwarf tidal system near NGC~3077, the Garland, to have the highest SFR
per unit luminosity among 150 galaxies of the Local volume with known SFRs.

  Being averaged over the local "cell of homogeneity" of 4 Mpc in diameter
around M~81, the rate of star formation of the group, $\dot{\rho}_{SFR}=
0.165 M_{\odot}/$year$\cdot$Mpc$^3$, proves to be 5--8 times higher
than, that of the average global rate at Z =0.

\end{abstract}

\bigskip

\section{Introduction}
Over the last years a great deal of observational effort has been undertaken
to understand how the star formation in a galaxy depends on its luminosity,
age and environment. One of the basic tools for this is measurement of
fluxes that are emitted by galaxies in the $H\alpha$  line.
Madau et al. (1996) have shown that the average star formation rate
has smoothly increased with the redshift from $Z\simeq5$ to $Z\simeq1$
and then has decreased steeply up to the present time (Z = 0).
In order to reliably fix the star formation density in a unit volume
at the present epoch, it is necessary to measure the $H\alpha$-fluxes
in a sufficiently representative sample of nearby galaxies limited by
a fixed distance. The most appropriate sample for this purpose is the
Catalog of 450 neighboring galaxies situated in a sphere of radius 10 Mpc
around the Milky Way (Karachentsev et al.\ 2004).

We have undertaken a comprehensive $H\alpha$ survey of these galaxies on the
northern sky with the BTA 6-m telescope. The first results have been published
for galaxies in the groups around NGC~6946 (Karachentsev et al.\ 2005)
and M~31 (Kaisin \& Karachentsev 2006). This program has also been
supported by observations of southern objects (Kaisin et al.\ 2006).
The data obtained allow the rates of  star formation in galaxies
to be studied in a uniquely wide range of luminosities, as high as
4 orders of magnitude.

In this paper, we present a complete summary of data on $H\alpha$-fluxes
for galaxies in a neighboring group around the giant spiral galaxy M~81.
An atlas of large-scale images of 20 dwarf galaxies of this group in the
B-band was presented by Karachentsev et al. (1985). Another 7 dwarf systems
have been discovered in this group ever since. According to
Karachentsev et al. (2002), about 30 galaxies are associated with M~81,
including the bright spiral galaxy NGC~2403 and five its satellites.
At present, accurate distances from the luminosity of the red giant branch
stars have been measured almost for all the members of M81, which has
made it possible to reconstruct the three-dimensional structure of the
complex. The loose clump of dwarf galaxies around NGC~2403 is situated at
the front boundary of the complex and is moving from us toward M81.
On the opposite more distant side there are the spiral NGC~4236 and few
irregular galaxies, which are not dynamically associated with M~81, but
take part in the cosmic expansion. On the whole, the entire complex of
these galaxies looks like a diffuse filament somewhat resembling
another nearby loose filament in Sculptor (Karachentsev 2005).

Unfortunately, no one has ever undertaken a systematic overview of
the M81 group population in the $H\alpha$ line. The brightest members
of the group were observed in $H\alpha$ by Hodge \& Kennicutt (1983),
Kennicutt et al. (1989), Miller \& Hodge (1994),
and Young et al. (1996). Later on, the $H\alpha$-fluxes
in several M81 companions were measured by Gil de Paz et al. (2003),
James et al. (2004), Hunter \& Elmegreen (2004), and
Lozinskaya et al. (2006). Miller (1996) has noted that the star formation
rates in the galaxies of the M81 group look, on the average, higher than
those in the Sculptor group and the Local group. However, about 2/3 of the
galaxies in the M81 group did not have measured $H\alpha$-fluxes
at that time. The survey of the dwarf population of this group,
which we have carried out, gives grounds for a correct comparison
of star formation processes in the most nearby groups.

\section{Observations and data reduction}

We obtained CCD images in the $H\alpha$ line and continuum for 36 members
of the M~81 group in the period from 2001 March to 2006 May with a median
seeing of $2.0\arcsec$. All the observations were made at the BTA 6-m
telescope of Special Astrophysical Observatory using the SCORPIO device
(Afanasiev at el.\ 2005) with a CCD chip of 2048$\times$2048 pixels,
and with a scale of 0.18$\arcsec$/pixel, which provides a total
field  of view of 6.1\arcmin$\times6.1\arcmin$. The images in
$H\alpha$+[NII] and continuum were obtained via observing the galaxies
though a narrow-band interference filter $H\alpha (\Delta\lambda=$75\AA\ )
with an effective wavelength $\lambda$=6555\AA\  and  medium-band
filters for the continuum: SED607 with $\lambda$=6063\AA\,
$\Delta\lambda$=167\AA, and SED707 with $\lambda$=7063\AA\,
$\Delta\lambda$=207\AA, respectively. Typical exposure times for most
the galaxies were $2\times300$s in the continuum and $2\times600$s in
$H\alpha$. Since the range of radial velocities in our sample is small, we
have used one and the same $H\alpha$ filter for all the observed objects.
The procedure of data reduction was standard for direct images obtained
with CCD. For all the data the bias was subtracted and then all the images
were flat-fielded, after which cosmic rays were removed and the sky
background was subtracted. The next operation was bringing into coicidence
all the images of a given object. Then all the images in the continuum were
normalized to the $H\alpha$  images with the use of 5--15 field stars and
subtracted. For the continuum subtracted images their $H\alpha$
fluxes were obtained,
using spectrophotometric standard stars observed
in the same nights as the objects. Investigation of measurement
errors has shown that they have typical values of 10--15\%.
\renewcommand{\baselinestretch}{1.0}
\begin{table}
\caption{The observational log}
\begin{tabular}{lclc} \\ \hline
Galaxy  &    Date    & Seeing  & $T_{exp}$ \\
\hline  &            &        &             \\
KKH34   & 13/11/2002 &  1.7"  & 1200s       \\
KKH37   & 09/10/2001 &  2.3"  & 1200       \\
DDO44   & 13/11/2002 &  1.8"  & 1200       \\
Holm II & 27/01/2004 &  2.1"  & 1200       \\
KDG52   & 05/12/2002 &  1.8"  & 1200:      \\
DDO53   & 28/11/2003 &  1.5"  & 1200       \\
U4483   & 28/11/2003 &  1.5"  & 1200       \\
VKN     & 05/12/2002 &  1.8"  & 1200       \\
HolmI   & 21/05/2006 &  2.3"  &  600:      \\
F8D1    & 01/02/2005 &  3.5"  & 1200       \\
FM1     & 01/02/2005 &  4.3"  & 1200       \\
N2976   & 03/11/2003 &  1.5"  &  600       \\
KK77    & 20/05/2006 &  2.2"  &  600       \\
BK3N    & 27/01/2004 &  2.0"  & 1200       \\
M82     & 23/05/2006 &  2.2"  & 1200:      \\
KDG61   & 21/05/2006 &  1.6"  &  600       \\
A0952+69& 27/01/2004 &  2.0"  & 1200       \\
KKH57   & 20/05/2006 &  1.8"  &  900       \\
N3077   & 03/11/2003 &  1.5"  & 1200       \\
Garland & 03/11/2003 &  1.5"  & 1200       \\
BK5N    & 02/02/2005 &  2.8"  & 1200       \\
KDG63   & 03/02/2005 &  1.8"  & 1200       \\
U5423   & 04/02/2006 &  2.1"  & 1200       \\
KDG64   & 03/02/2005 &  1.7"  & 1200       \\
IKN     & 05/12/2002 &  1.2"  & 1200:      \\
HIJASS  & 29/01/2004 &  1.9"  &  600       \\
HS117   & 29/01/2004 &  2.5"  & 1200       \\
DDO78   & 04/02/2005 &  1.7"  &  600       \\
I2574   & 04/02/2006 &  4.4"  & 1200:      \\
DDO82   & 07/02/2003 &  1.9"  & 1200       \\
BK6N    & 21/03/2006 &  4.1"  & 1200       \\
DDO87   & 21/05/2006 &  1.7"  &  600       \\
KDG73   & 07/02/2003 &  2.1"  & 1200       \\
U6456   & 07/02/2003 &  2.2"  & 1200       \\
U7242   & 20/05/2006 &  2.7"  &  300       \\
N4605   & 29/01/2004 &  2.0"  & 1200       \\
DDO165  & 17/03/2001 &  3.9"  &  900       \\
\hline
\end{tabular}
\end{table}

The conditions of our observations are given in Table 1. Its columns
indicate (1) the galaxy name, (2) the date of observations, (3) the
average seeing (FWHM), and (4) the total exposure time in seconds;
here, colon means that the sky was not photometric.

\section{Results}

Fig.1 displays (from left to right) the images of [$H\alpha$ plus
continuum] and [$H\alpha$ minus continuum] for 36 observed galaxies.
The size of the presented frames is $4\arcmin\times4\arcmin$,
the north-east directions are marked by the arrows, the names of
the galaxies are given in the upper right corner of the frames.
For two large galaxies: M82 and IC2574, we present a mosaic of images
of three frames, these are at the end of Fig.1.
Some basic parameters of the galaxies that we have observed are given
in Table 2. To make the picture complete, we have also added into
the table a few bright members of the group whose fluxes in the
$H\alpha$ line  were measured earlier by other authors. The columns of
Table 2 contain  the following data taken, as a rule, from the CNG catalog
(Karachentsev et al.\ 2004): (1) the galaxy name; (2) equatorial
coordinates for the epoch J2000.0; (3) the distance to a galaxy in Mpc
with allowance made for new measurements (Karachentsev et al.\ 2006);
(4) the blue absolute magnitude of a galaxy with the given distance after
correction for the Galactic extinction $A_b$ from Shlegel et al. (1998);
(5) the major linear diameter in kiloparsecs corrected for the galaxy
inclination and the Galactic extinction in the manner adopted by
de Vaucouleurs, de Vaucouleurs, \& Corwin (1976); (6) morphological type;
(7) ``tidal index'' (TI) following from the CNG: for every galaxy
``$i$'' we have found its ``main disturber''(MD), producing the
highest tidal action
$$TI_i = \max  \{\log(M_k/D_{ik}^3)\} + C,\;\;\;(i = 1, 2...  N)$$
where  $M_k$ is the total mass of any neighboring potential MD galaxy
(proportional to its luminosity with $M/L_B = 10 M_{\odot}/L_{\odot}$)
separated from the considered galaxy by a space distance $D_{ik}$;
the value of the constant $C$  is chosen so that $TI = 0$ when the
Keplerian cyclic period of a galaxy with respect to its MD equals
the cosmic Hubble time, $T_{0}$; therefore positive values correspond
to galaxies in groups, while the negative ones correspond to field
galaxies; (8) the logarithm of the hydrogen mass of a galaxy,
$\log(M_{HI}/M_{\odot})= \log F_{HI}+2\log D_{Mpc}+5.37$,
defined from its flux $ F_{HI}$ in the 21 cm line; in some dwarf spheroidal
galaxies the upper limit of the flux was estimated from the observations
by  Huchtmeier et al. (2000); (9) the integral flux of the galaxy in
the $H\alpha$ +[NII] lines expressed in terms of $10^{-16}$erg/cm$^2$ sec
with indication of a typical  measurement error; the asteriscs denote
data sources on SFR according to other authors, reduced to the distance of
the galaxy adopted in Table 2;(10) the star formation rate in
the galaxy on a logarithmic scale, SFR($M_{\odot}$/year) =
1.27$\cdot 10^9 F_c(H_{\alpha})\cdot D^2$ (Gallacher et al.\ 1984),
where the integral flux in the $H\alpha$ line is corrected for
extinction as $A(H\alpha) = 0.538\cdot A_b$, while the galaxy
distance is expressed in Mpc; (11,12) the dimensionless parameters
$p_*=\log([SFR]\cdot T_0/L_B)$ and $f_*=\log(M_{HI}/[SFR]\cdot T_0)$,
which characterize the past and the future of the process of star formation;
here $L_B$ denotes the total blue luminosity of the galaxy in units of solar
luminosity, while $T_0$ is the age of the universe assumed equal to 13.7
billion years (Spergel et al.\ 2003); the last column contains the indication
of the data sources on SFR according to other authors, reduced to the
distance of the galaxy adopted in Table 2.

Below we note some features of emission regions in the galaxies
that we observed.

{\em KKH 34}. This dwarf irregular galaxy at the fartherst outskirts of
the group M81 (from the side of the neighboring group IC342/Maffei)
shows faint diffuse emission in its central part.

{\em KKH~37}. This is another dIr galaxy located half-way between M81
and IC~342. Apart from the diffuse $H\alpha$ emission, KKH~37 shows
the presence of a compact HII region near the center, which has
been used by Makarov et al. (2003) to determine the radial velocity
of the galaxy.

{\em DDO~44}. This dwarf spheroidal (dSph) companion of the spiral galaxy
NGC~2403 does not show emission in the neutral hydrogen line HI. We have not
detected $H\alpha$ emission within the optical boundary of DDO~44. However,
there is a compact patch (marked by a circle in the figure) at the NE edge
of the galaxy. It may be rather a peculiar background galaxy than an
emission knot belonging to DDO~44. The $H\alpha$ flux presented in Table 2
refers exactly to it. Apparently, to establish its nature, spectral
observations are needed.

{\em HolmII}. The irregular galaxy with powerful star formation sites, whose
periphery is beyong our frame. The $H\alpha$ flux presented in
Table 2 takes into account the correction for the incomplete field
$(\Delta\log F=0.20)$ under assumption that distribution of the $H\alpha$
emission is proportional to the galaxy blue surface brightness.

\renewcommand{\baselinestretch}{0.8}
\begin{deluxetable}{lcp{0.5cm}rrp{0.2cm}rrp{1.5cm}rrr}
\footnotesize
\tablecaption{M81 group: SFR and M(HI) estimates}
\tablewidth{0pt}
\tablehead{
\colhead{Name}&
\colhead{RA(2000.0)Dec}&
\colhead{D}&
\colhead{$M_B$}&
\colhead{$A_{25}$}&
\colhead{Type}&
\colhead{TI}&
\colhead{LgMHI}&
\colhead{Flux}&
\colhead{LgSFR}&
\colhead{p*}&
\colhead{f*}
\\
& &\colhead{Mpc}&
\colhead{mag}&
\colhead{kpc}&
& &
\colhead{M$_{\sun}$}&
&\colhead{M$_{\sun}$/yr}&
& \\
\colhead{(1)}&
\colhead{(2)}&
\colhead{(3)}&
\colhead{(4)}&
\colhead{(5)}&
\colhead{(6)}&
\colhead{(7)}&
\colhead{(8)}&
\colhead{(9)}&
\colhead{(10)}&
\colhead{(11)}&
\colhead{(12)}
}
\startdata
KKH 34& 055941.2+732539 &4.61 &$-$12.30&  1.34 & Ir   & $-$1.8& 7.08 &     36$\pm$3     &$-$3.78 &$-$0.72   & $ $0.72  \\
KKH 37& 064745.8+800726 &3.39 &$-$11.59&  1.17 & Ir   &  1.2&  6.66 &    110$\pm$13    &$-$3.73 &$-$0.39    & $ $0.25  \\
N 2366& 072852.0+691219 &3.19 &$-$16.02&  5.71 & IBm  &  1.0&  8.85 &       a)      &$-$0.85 &$ $0.72       & $-$0.44  \\
DDO 44& 073411.3+665310 &3.19 &$-$12.07&  2.67 & dSph &  1.7& $<$6.0  &      6$\pm$6     &$-$5.07 &$-$1.92: & $ $0.93: \\
N 2403& 073654.4+653558 &3.30 &$-$19.29& 19.43 & Scd  & $-$0.0&  9.52 &       b)      &$-$0.04 &$ $0.22     & $-$0.58  \\
HolmII& 081905.9+704251 &3.39 &$-$16.72&  7.78 & Im   &  0.6&  8.99 &  56950$\pm$330   &$-$1.05 &$ $0.24    & $-$0.10  \\
KDG 52& 082356.0+710146 &3.55 &$-$11.49&  1.35 & Ir   &  0.7&  7.12 &      5$\pm$5     &$-$5.08 &$-$1.70:   & $ $2.06  \\
DDO 53& 083406.5+661045 &3.56 &$-$13.37&  1.67 & Ir   &  0.7&  7.61 &   5310$\pm$41    &$-$2.04 &$ $0.59    & $-$0.49  \\
U 4483& 083703.0+694631 &3.21 &$-$12.73&  1.04 & BCD  &  0.5&  7.51 &   3168$\pm$29    &$-$2.35 &$ $0.54    & $-$0.28  \\
VKN   & 084008.9+682623 &     &       &       &      &     &       &     12$\pm$9      & $-$4.6: & $-$      &  $-$  \\
HolmI & 094028.2+711111 &3.84 &$-$14.49&  4.03 & Ir   &  1.5&  8.13 &   3565$\pm$31    &$-$2.13 &$ $0.05    & $ $0.12  \\
F8D1  & 094450.0+672832 &3.77 &$-$12.59&  2.58 & dSph &  2.0& $<$6.2  &     24$\pm$4     &$-$3.86 &$-$0.92: & $-$0.08: \\
FM1   & 094510.0+684554 &3.42 &$-$10.48&  0.93 & dSph &  1.8& $<$6.1  &      7$\pm$6     &$-$4.92 &$-$1.13: & $ $0.88: \\
N 2976& 094715.6+675449 &3.56 &$-$17.10&  5.57 & Sdm  &  2.7&  8.27 &  57870$\pm$300   &$-$0.97 &$ $0.17    & $-$0.90  \\
KK 77 & 095010.0+673024 &3.48 &$-$12.03&  2.61 & dSph &  2.0& $<$6.1  &      6$\pm$2     &$-$4.90 &$-$1.73: & $ $0.86: \\
BK3N  & 095348.5+685809 &4.02 &$-$9.59&  0.60 & Ir   &  1.0& $<$6.5  &      3$\pm$2     &$-$5.14 &$-$1.00:  & $ $1.50: \\
M 81  & 095533.5+690360 &3.63 &$-$21.06& 26.85 & Sab  &  2.2&  9.46 &       c)      &$-$0.12 &$-$0.56       & $-$0.56  \\
M 82  & 095553.9+694057 &3.53 &$-$19.63& 10.93 & I0   &  2.7&  8.90 &       d)      &$ $0.42 &$ $0.55       & $-$1.66  \\
KDG 61& 095702.7+683530 &3.60 &$-$12.85&  2.40 & dSph?&  3.9& $<$6.2  &    439$\pm$13    &$-$3.08 &$-$0.24  & $-$0.86: \\
A0952 & 095729.0+691620 &3.87 &$-$11.51&  2.14 & Ir   &  1.9&  7.0  &    351$\pm$5     &$-$3.10 &$ $0.28    & $-$0.04  \\
HolmIX& 095732.4+690235 &3.7  &$-$13.68&  2.77 & Ir   &  3.3&  8.50 &       e)      &$-$2.65 &$-$0.14       & $ $1.01  \\
KKH 57& 100016.0+631106 &3.93 &$-$10.19&  0.67 & dSph &  0.7& $<$6.3  &      7$\pm$6     &$-$4.85 &$-$0.95: & $ $1.01: \\
N 3077& 100321.0+684402 &3.82 &$-$17.76&  6.14 & I0p  &  1.9&  8.79 &  52950$\pm$180   &$-$0.95 &$-$0.07    & $-$0.40  \\
Garlnd& 100342.0+684136 &3.82 &$-$11.40&  6.60 & Ir   &  4.0&  7.54 &   4210$\pm$390   &$-$2.05 &$ $1.37    & $-$0.55  \\
BK5N  & 100440.3+681520 &3.78 &$-$10.61&  0.87 & dSph &  2.4& $<$6.4  &      5$\pm$4     &$-$4.99 &$-$1.25: & $ $1.25: \\
KDG 63& 100507.3+663318 &3.50 &$-$12.12&  1.84 & dSph?&  1.8& $<$6.3  &     47$\pm$3     &$-$4.05 &$-$0.92  & $ $0.21: \\
U 5423& 100530.6+702152 &5.3  &$-$14.54&  1.37 & BCD? & $-$0.9&  7.40 &   1601$\pm$23    &$-$2.17 &$-$0.01  & $-$0.57  \\
KDG 64& 100701.9+674939 &3.70 &$-$12.57&  1.85 & dSph?&  2.5& $<$6.3  &     24$\pm$5     &$-$4.33 &$-$1.38  & $ $0.49: \\
IKN   & 100805.9+682357 &3.75 &$-$12.13&  3.19 & dSph &  2.7& $<$6.3  &      8$\pm$5     &$-$4.79 &$-$1.66: & $ $0.95: \\
HIJASS& 102100.2+684160 &3.7  &$-$7.93&       & HIcld&  2.2&  8.18 &     54$\pm$3     &$-$4.01 &$ $0.80:    & $  $2.05 \\
HS 117& 102125.2+710658 &3.96 &$-$11.98&  1.72 & dSph &  1.0& $<$5.0  &      8$\pm$2     &$-$4.70 &$-$1.51: & $-$0.44: \\
DDO 78& 102627.9+673924 &3.72 &$-$12.17&  2.20 & dSph &  1.8& $<$6.2  &      7$\pm$7     &$-$4.89 &$-$1.78: & $ $0.95: \\
I 2574& 102822.4+682458 &4.02 &$-$17.46& 13.35 & Im   &  0.9&  9.23 &       f)      &$-$0.62 &$ $0.38       & $-$0.29  \\
DDO 82& 103035.0+703710 &4.00 &$-$14.63&  3.70 & Im   &  0.9& $<$6.8  &   1717$\pm$34    &$-$2.42 &$-$0.29  & $-$0.92: \\
BK6N  & 103431.9+660042 &3.85 &$-$11.08&  1.14 & dSph &  1.1& $<$6.3  &      7$\pm$3     &$-$4.87 &$-$1.32: & $ $1.03: \\
DDO 87& 104936.5+653150 &7.4  &$-$14.42&  5.11 & Ir   & $-$1.5&  8.03 &    603$\pm$8     &$-$2.37 &$-$0.16  & $ $0.26  \\
KDG 73& 105255.3+693245 &3.70 &$-$10.83&  0.63 & Ir   &  1.3&  6.51 &     13$\pm$4     &$-$4.63 &$-$0.98    & $ $1.00  \\
U 6456& 112800.6+785929 &4.34 &$-$14.03&  1.72 & BCD  & $-$0.3&  7.79 &   7343$\pm$62    &$-$1.72 &$ $0.65  & $-$0.63  \\
U 7242& 121407.4+660532 &5.42 &$-$14.15&  2.56 & Ir   & $-$0.5&  7.70 &    597$\pm$230   &$-$2.64 &$-$0.32  & $ $0.20  \\
N 4236& 121643.3+692756 &4.45 &$-$18.59& 23.58 & Sdm  & $-$0.4&  9.46 &       g)      &$-$0.65 &$-$0.11     &  $-$0.03 \\
N 4605& 124000.3+613629 &5.47 &$-$18.07&  7.79 & Sdm  & $-$1.2&  8.54 &  65770$\pm$160   &$-$0.59 &$ $0.16  & $-$1.01  \\
DDO165& 130626.8+674215 &4.57 &$-$15.09&  4.20 & Im   &  0.0&  8.14 &   1121$\pm$36    &$-$2.51 &$-$0.57    & $ $0.51  \\
\enddata
\tablenotetext{a}{ N 2366, James et al. 2004;}
\tablenotetext{b}{ N 2403, Kennicutt et al. 1989;}
\tablenotetext{c}{ M 81, Kennicutt et al. 1989;}
\tablenotetext{d}{ M 82, Young et al. 1996;}
\tablenotetext{e}{ HolmIX, James et al. 2004;}
\tablenotetext{f}{ I 2574, Miller \& Hodge, 1994;}
\tablenotetext{g}{ N 4236, Kennicutt et al. 1989.}
\end{deluxetable}
\protect

{\em KDG~52}. The detailed distribution of neutral hydrogen and the radial
velocity field in this dIr galaxy has been obtained by Begum et al. (2006).
Judging by the images taken with the HST (Karachentsev et al.\ 2002),
KDG~52 contains a young blue stellar population. However, Fig.1 does not show
any noticeable $H\alpha$ emission above a detection limit of
$5\cdot10^{-16}$ erg/cm$^2$sec.

{\em DDO~53 and UGC~4483}. These are known dIr and BCD galaxies with
bright HII regions (Hunter et al.\ 2004, Gil de Paz et al.\ 2003).

{\em VKN}. This is an object of extremely low surface brightness in the
vicinity of UGC~4483, which is likely to be a part of faint Galactic cirrus.

{\em HolmI}. The irregular galaxy of low surface brightness with a lot
of compact and diffuse emission knots on its periphery.

{\em F8D1, FM1 and KK~77}. These are dwarf spheroidal companions to M81 of
very low surface brightness. In the vicinity of F8D1 one can see three faint
knots (marked by circles), whose possible emission nature needs special
check. The $H\alpha$ flux for F8D1 given in Table 2 refers to these knots.

{\em NGC~2976}. This is a late-type Sdm galaxy with unclear spiral
structure outlined by bright sites of star formation and dust complexes
(James et al.\ 2004).

{\em BK3N}. The faintest of the known M81 companions. Judging by
the images obtained with the HST, it has a population of young stars but
does not show signs of the $H\alpha$ emission above the the detection limit
of $2\cdot10^{-16}$ erg/cm$^2$sec. The value of $HI$- flux in BK3N is
also rather uncertain because of the projected HI envelope of M81.

{\em KDG~61}. The compact emission region discovered by Johnson et al. (1997)
on the northern side of this galaxy is not typical of dwarf spheroidal
objects. A very faint filament structure is also seen in Fig.1 around this
compact bright HII region.

{\em A~0952+69}. This is a part of the ``Arp loop'' resolved  into
stars by means of WFPC2 at the HST (Karachentsev et al.\ 2002).
The presence of numerous small HII regions is evidence of the star
formation process in this peculiar structure at the outskirts of M81,
occuring at low rates.

{\em KKH~57 and BK5N}. Spheroidal dwarf systems without signs of star
formation.

{\em NGC~3077 and Garland}. The $H\alpha$ emission in NGC~3077 shows strong
concentration towards the galaxy center, where dust clouds also occured.
Closer to the outskirts the emission has a filament structure looking like
crab's claws. At the southern periphery of NGC~3077 a chain of
associations of blue stars and HII regions is seen, which was called
``Carland'' (Karachentseva et al.\ 1985,  Karachentsev  et al.\ 1985).
With this unusual structure which is likely to be of tidal origin
(Makarova et al.\ 2003), neutral hydrogen filament and molecular clouds
are associated ( Yun et al. 1994, Walter et al.\ 2002).

{\em KDG~63 and KDG~64}. Both galaxies are known as dSph systems with an
old stellar population. However, on our $H\alpha$ images of KDG~63 and KDG~64
two and one knots are seen, respectively, (marked by circles in Fig.1),
which may prove to be compact HII-regions.

{\em  UGC~5423}. Judging by the radial velocity , $V_{LG}=+496$km s$^{-1}$,
and the distance, 5.3 Mpc, this BCD/dIr galaxy with emission knots is
situated behind the group M81 or else at its farthest outskirts.

{\em IKN}. This very diffuse galaxy is barely seen in a halo of a
bright star neighboring from the north. The galaxy $H\alpha$ emission
is visibly absent.

{\em HIJASS}. This intergalactic hydrogen cloud has been detected by Boyce
et al. (2001) and investigated in detail at the VLA by Walter et al. (2005).
Its optical counterpart, if exists, has a total apparent magnitude
of about 20$^m$. In the region of this object, one can see in Fig.1
a few faint diffuse knots whose nature can be established by
future spectral observations.

{\em HS~117}. Basing on its HST images (Karachentsev et al.\ 2006),
this galaxy of regular shape contains a small number of bluish stars.
However, the observations of it with GMRT (Begum, 2006), has not
revealed sings of the HI emission. On the $H\alpha$ image of HS~117 near
the center, there is a compact knot (marked by a circle), a spectrum
of which we are going to obtain soon.

{\em DDO~78 and BK6N}. Two dSph systems in each of which we suspect
one emission knot (marked by circles). However, judging by the HST
images, these are artefacts caused by incomplete subtraction of
images of distant galaxies. The $H\alpha$ fluxes for them
given in Table 2 refer to these knots.

{\em  DDO~82 and UGC~7242}. Irregular galaxies whose images are strongly
contaminated by neighboring bright stars. A considerable $H\alpha$
emission is visible in both galaxies.

{\em DDO~87}. This dIr galaxy with a radial velocity $V_{LG}=+468$km s$^{-1}$
and a distance of 7.4 Mpc (estimated from the brightest blue stars) is
likely to be located behind the group M81. About a dozen compact HII-regions
are seen in the galaxy body with practically complete absence of diffuse
emission.

{\em KGD~73}. This is a dIr galaxy of low surface brightness. The
color-magnitude diagram for it obtained with WFPC2 of HST shows the
presence of blue stars. In the optical domain of the galaxy we note two
very faint diffuse emission regions whose fluxes are at the limit of
detection.

{\em UGC~6456 = VIIZw403}. Along with UGC~4483 and UGC~5423 this is a third
blue compact dwarf galaxy (BCD) in our sample. The bright regions of star
formation in it are located assymetrically to the east with respect to the
geometrical center of the galaxy.

{\em NGC~4605}. It is surprising, but this bright galaxy has never been
imaged before in the $H\alpha$ line. A powerful emission is observed in its
disk, which exceeds in flux the $H\alpha$ emission from another Sdm galaxy,
NGC~2976. A system of fine emission filaments attaching to the emission
disk of NGC~4605 makes the galaxy looking like a bristled up sea-horse.
A long diffuse filament extends on the southern side of the disk while
the western side of the disk is likely to contain a great amount of dust.

{\em DDO~165}. An irregular galaxy on the far side of the group M81.
There are several compact HII-regions in its optical domain along
with the common emission envelope.

As we have already noted, the $H\alpha$ fluxes in a number of galaxies
in the M81 group were measured earlier by other authors. We present in
Table 3 the data on the star formation rates in 12 members of the group
obtained from our measurements (6m) as well as from estimates by other
authors reduced to the distances indicated in Table 2. As can be seen,
the agreement of log[SRF] proves to be quite satisfactory. The exception is
only the discrepancy with the data by Walter et al. (2002, 2006) for
NGC~3077 and Garland, the cause of which is not clear to us. For the rest
of the galaxies, the average difference of our and other estimates makes
$<\Delta$log[SFR]$>=-0.02\pm0.02$ with a typical external error of a single
measurement of the $H\alpha$ flux of about 10\%.

The $H\alpha$ line images of galaxies in the M81 group obtained by us
and other authors show that the star formation processes in the galaxies
belonging to one and the same group are characterized by great variety. In
some relatively luminous galaxies (NGC~3077, M82) the main emission comes
from the galaxy core. In other bright spiral galaxies (NGC~2976, NGC~4605,
NGC~2403, and M~81) the $H\alpha$ emission is distributed more or less
uniformly, over the whole disk. In irregular galaxies with a typical
luminosity of the Magellanic clouds (HolmII, IC~2574, NGC~2366) one
observes the presence of powerful sites of star formation, and rather often
such superassociations being located at the outskirts of these
galaxies. In dIr and BCD galaxies of low luminosity, small compact
HII-regions (HolmI, DDO~87, UGC~4483, DDO~165) or else separate diffuse
emission knots (KK~34, KKH~37) are characteristic features. The presence
of emission knots in dwarf galaxies, which have been classified as
spheroidal (KDG~61, KDG~63, HS~117) proved to be unexpected, although some
of the detected emisssion details, need an additional spectral confirmation.
At least, we note the cases where dIr galaxies of very low $(-10,-11^m$)
luminosity: KDG~52 and BK3N do not show any detectable $H\alpha$
flux, although they contain a blue stellar population, seen on the
images taken with the HST. Apparently, the potential well in these pygmy
galaxies is so shallow that it is unable to hold ionized gas.

\begin{table}
\footnotesize
\caption{Comparison of [SFR] estimates for the M81 group galaxies}
\begin{tabular}{p{1.0cm}p{1.2cm}p{1.2cm}p{1.cm}l} \\ \hline
Galaxy   &  log[SFR]& log[SFR]&  $\Delta$   &      source           \\
  &    6m        &    oth        &          &                        \\
\hline
 HolmII  &  -1.05     &   -1.01    & -0.04    &  Hunter et al. 2004    \\
 KDG52   &  -5.08     & $<$-4.64    &   -      & Miller \& Hodge 1994                  \\
 DDO53   &  -2.04     &   -2.04    &  0.00    &  Hunter et al. 2004    \\
 U4483   &  -2.35     &   -2.33    & -0.02    &  Gil de Paz et al. 2003\\
 HolmI   &  -2.13     &   -2.09    & -0.04    &  James et al. 2004     \\
 N2976   &  -0.97     &   -0.93    & -0.04    &  James et al. 2004     \\
 N3077   &  -0.95     &   -0.86    & -0.09    &  James et al. 2004     \\
 U5423   &  -2.17     &   -2.21    &  0.04    &  van Zee 2000          \\
 DDO82   &  -2.42     &   -2.40    & -0.02    &  James et al. 2004     \\
 DDO87   &  -2.37     &   -2.30    & -0.07    &  James et al. 2004     \\
 U6456   &  -1.72     &   -1.86    &  0.14    &  Gil de Paz et al. 2003 \\
 DDO165  &  -2.51     &   -2.44    & -0.07    &  James et al. 2004       \\
\hline                                                                  \\
 N3077   &  -0.95     &   -1.15    &  0.20    &  Walter et al. 2002      \\
 Garland &  -2.05     &   -2.53    &  0.48    &  Walter et al. 2006      \\
\hline
\end{tabular}
\end{table}

\setcounter{figure}{1}
\begin{figure}[h]
\includegraphics[angle=-270,scale=.3]{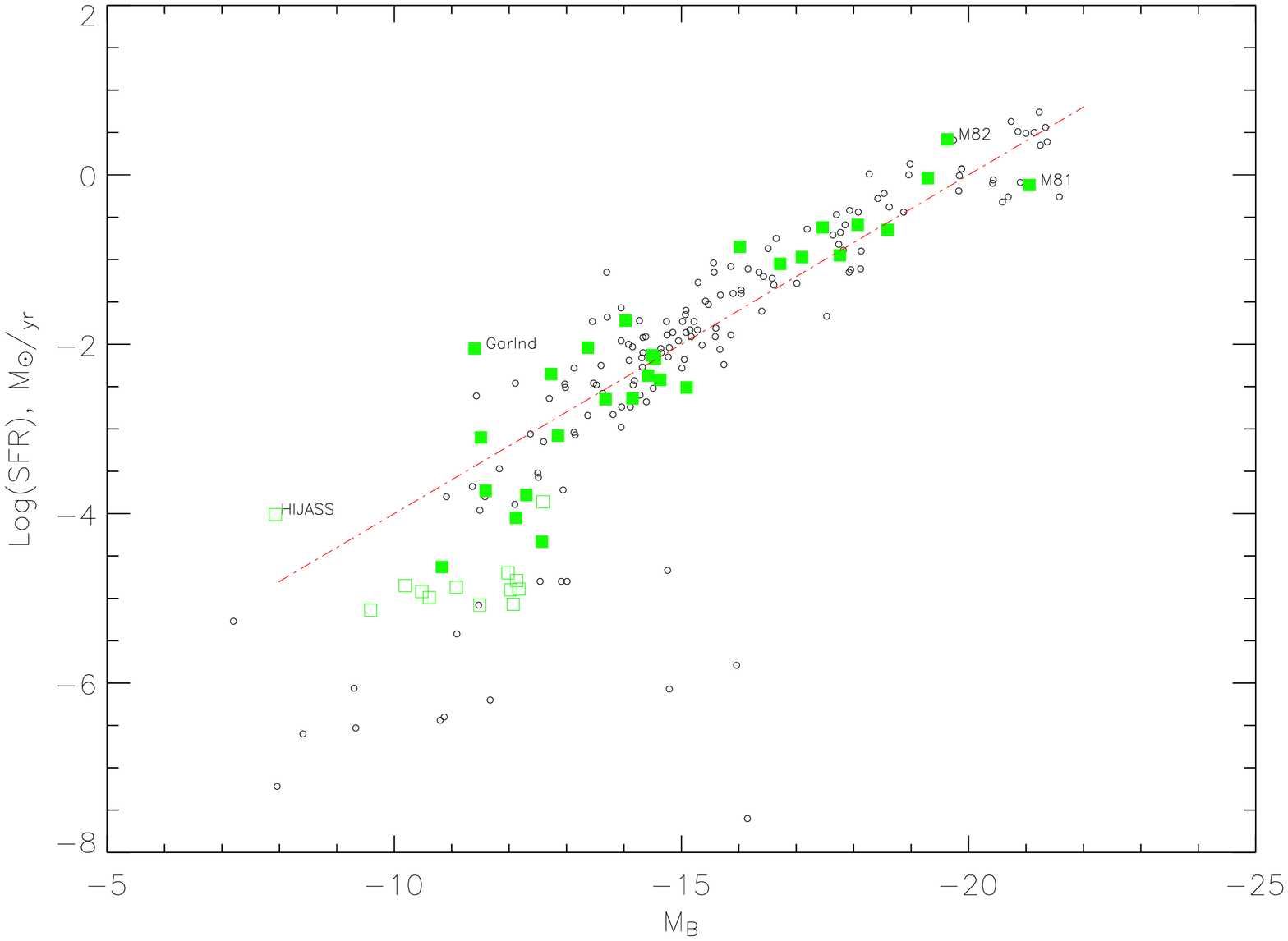}
\caption{Star formation rate versus blue absolute magnitide for 150 nearby
galaxies of the Local volume (open circles). Members of the M81 group are
shown by squares. The open squares indicate the M81 dwarf companions
with only upper limit of their SFR. The straight line correspondes to a
constant SFR per unit luminosity.}
\end{figure}

The distribution of 41 galaxies in the region of M81 versus their
absolute magnitude $M_B$ and SFR is presented in Fig.2 by squares.
The dwarf galaxies with only the upper limit of the $H\alpha$
flux are indicated by open squares. For comparison, we also present
another 150 galaxies from the Local volume ($D<10$ Mpc) shown by small
circles on this diagram. The SFR data for them are taken from
Hodge \& Kennicutt (1983), Kennicutt et al. (1989), Miller \& Hodge (1994),
Young et al. (1996), Gil de Paz et al. (2003), James et al. (2004),
Hunter \& Elmegreen (2004), Karachentsev et al.\ (2005), and
Kaisin \& Karachentsev (2006). Here, objects with a low star formation
rate  [SFR]$ < 3\cdot10^{-6}M_{\odot}$/year all are companions to the
neighboring spiral M31.

As can be seen from this "giraffe-like" diagram, most of the galaxies
with magnitudes from $-13^m$ to $-21^m$, situated along the giraffe's
neck, follow a linear relationship [SFR]$\propto L_B$ (shown by the
dashed line) with a r.m.s. scatter $\sigma(\log$[SFR]$)\simeq0.5$.
The M81 group members on this diagram follow the common relation without
significant displacement with respect to other Local volume galaxies.

Among the galaxies of the M81 suite, Garland is distinguished by the
highest star formation rate per unit luminosity (i.e. specific SFR).
Apparently, we observe this tidal structure at the peak of star formation
process, in a burst. Among other 150 galaxies of the Local volume,
only one BCD galaxy, UGCA~281 = Mkn~209, approaches the Garland in its
very high specific SFR.

\setcounter{figure}{2}
\begin{figure}[h]
\includegraphics[angle=0,scale=.5]{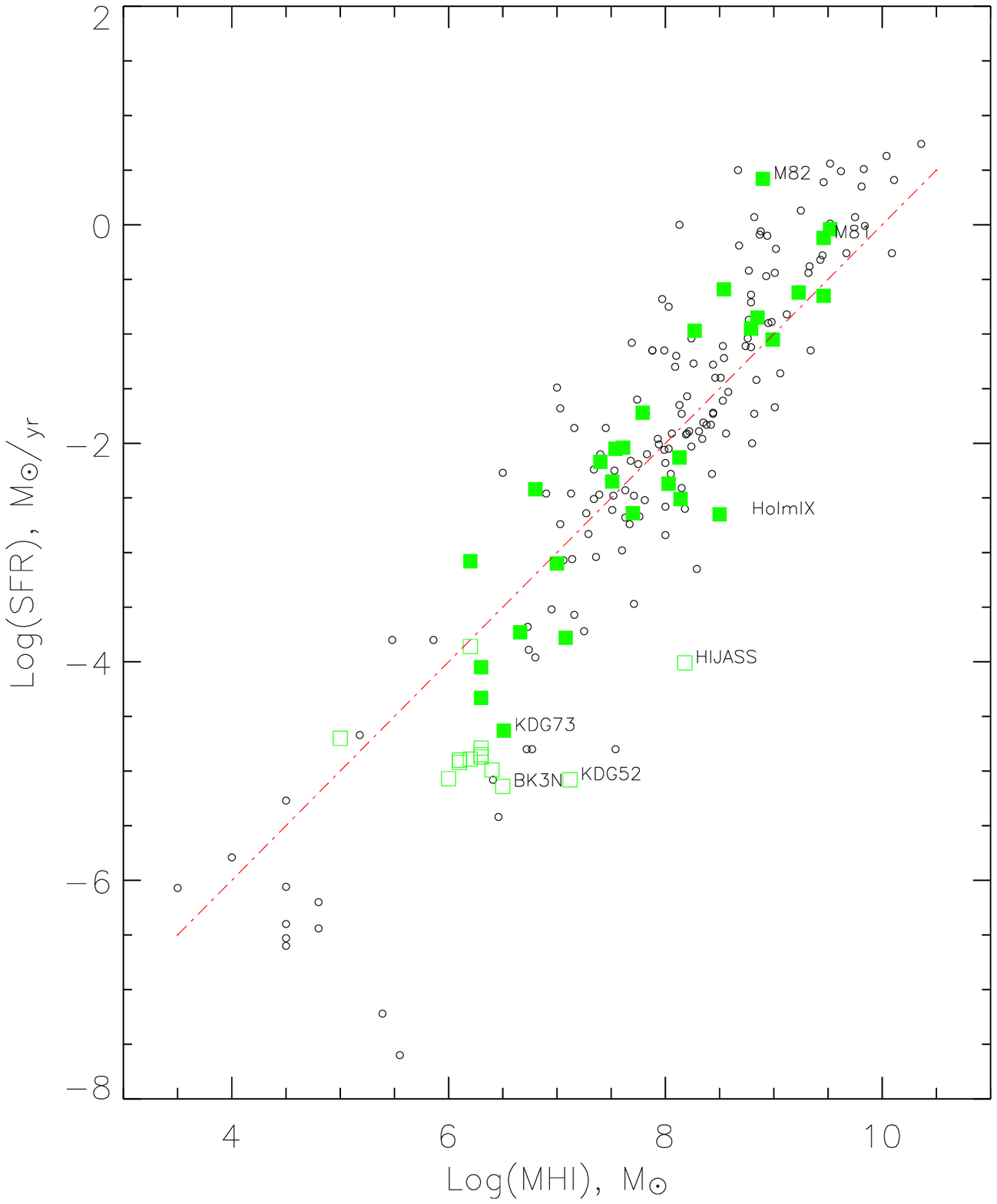}
\caption{Star formation rate versus neutral hydrogen mass for nearby
galaxies situated within 10 Mpc (open circles). Members of the
M81 group are shown by squares. The group members with upper limit of
SFR or $M_{HI}$ are indicated by open squares. The
straight line correspondes to a fixed SFR per unit hydrogen mass.}
\end{figure}

Another useful diagram, showing the evolutionary status of a galaxy, is
presented in Fig.3, where the global star formation rate of galaxies is
compared with the amount of mass of neutral hydrogen in them. The
designations of the M81 group members and other galaxies of the Local volume
are given here the same as in the preceding figure. Many authors:
Kennicutt (1989, 1998), Taylor \& Webster (2005), Taylor (2005),
Juneau et al. (2005), Tutukov (2006), Feulner et al. (2006),
and  Gutierrez et al. (2006), have been concerned with the
interpretaion of the character of distribution of galaxies over the
diagram [SFR] $\propto\ M_{HI}$. As has been noted by
these authors, spiral and irregular galaxies follow, on the average,
the relation [SFR] $\propto{M}^{1.4}_{HI}$, that is, a higher star
formation rate is observed in galaxies with larger hydrogen masses.
In other words, the use up of the present reserve of gas to from a
stellar component occurs in dwarf galaxies in a retarded (``letargic'')
mode in comparison with spiral ones, which favours long life of dwarf
galaxies. An extreme example here is KDG~52 where there is a young stellar
population, but the $H\alpha$ flux is not amenable to measuring at usual
(20 minutes) exposure times. An attempt to detect the $H\alpha$ emission
in KDG~52 and BK3N at much longer exposures seems to be of interest.
It should be noted here that from the data by Appleton et al. (1981)
and Yun et al. (1994), the central region of the M81 group is filled
with HI clouds and filaments, which makes the estimates of the HI mass
in BK3N, HoIX, A0952+69, and Garland rather uncertain.

As we have already noted, the suitable parameters to describe the past
and future star formation process in a galaxy are dimensionless quantities:
$p_*$=log$([SFR]\cdot T_0/L_B)$ and $f_*$= log$(M_{HI}/[SFR]\cdot T_0)$.
The former characterizes what proportion of its luminosity the galaxy would
produce during the Hubble time $T_0$ at the current rate of star formation
and the mass-to-luminosity ratio $1 M_{\odot}/L_{\odot}$. The latter
parameter shows how much the Hubble time the galaxy will need to spend
the present supply of gas if star formation proceeds at the
currently observed rate. The distribution of all members of
the M81 group on the plane $\{f_*, p_*\}$ is displayed in Fig.4.
The open squares correspond to the objects in which only the upper
limit of the $H\alpha$ flux or of the hydrogen mass was measured.

\setcounter{figure}{3}
\begin{figure}[ht]
\includegraphics[angle=0,scale=.5]{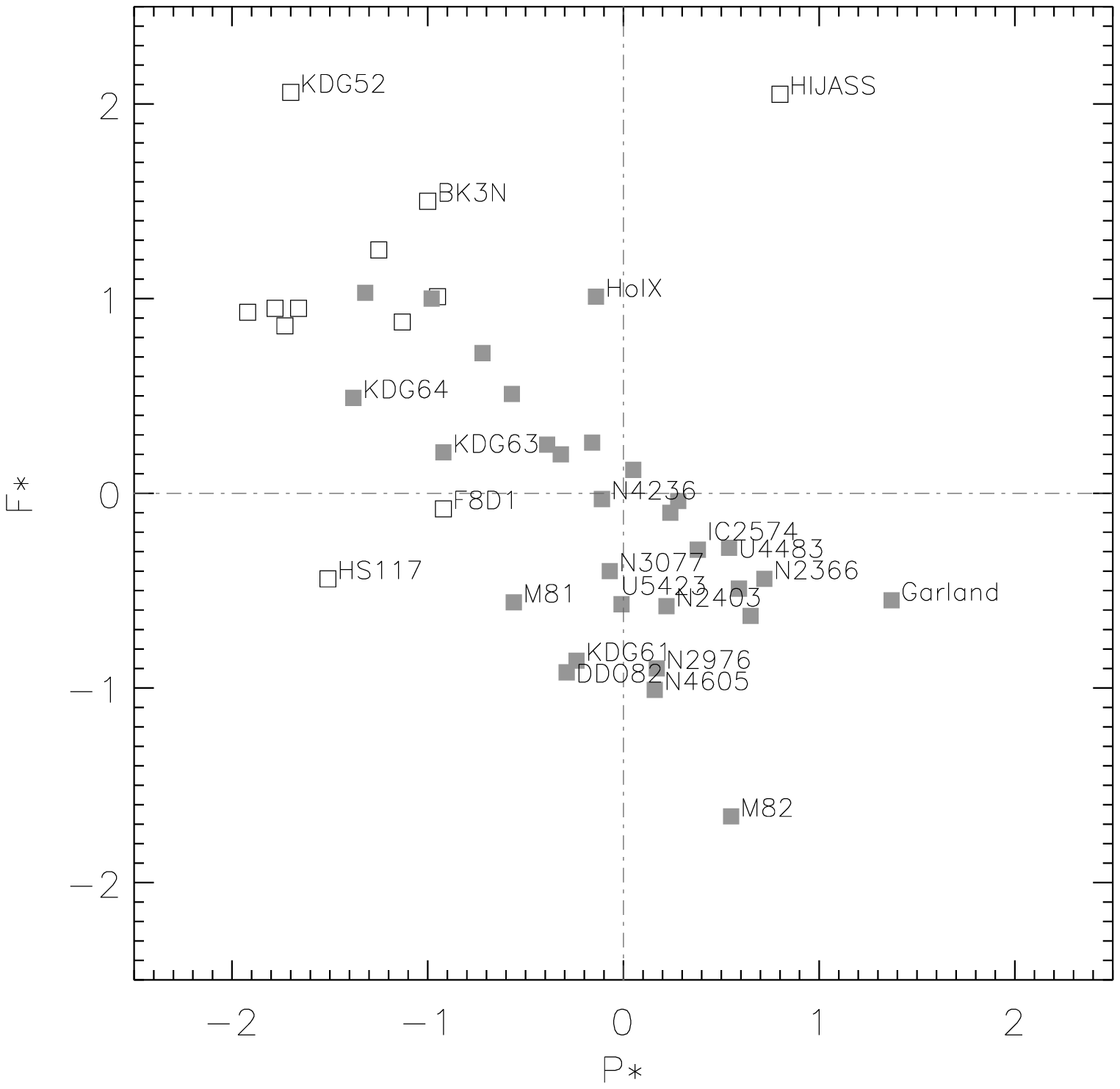}
\caption{The M81 group galaxies on the evolution plane "past-future":
$p_*$=log$([SFR]\cdot T_0/L_B)$ and $f_*$= log$(M_{HI}/[SFR]\cdot T_0)$.
The galaxies with upper limits of SFR or $M_{HI}$ are shown by open
squares.}
\end{figure}

  The average star formation indicators for the M81 group members of
different morphological types are given in Table 4. Its last row indicates
the mean specific star formation rate attributed to 1 $(kpc)^2$.
The following conclusions can be drawn from the data obtained for the M~81
group galaxies.

\begin{table}
\footnotesize
\caption{Mean star formation indicators for different morphological types}
\begin{tabular}{ccccc} \\ \hline
Parameters       &     Sa-m     &  Im,BCD    &    dIr     &    dSph    \\
\hline
 Number          &       5      &    10      &    12      &     13     \\
		 &              &            &            &            \\
 $M_B$           &    $-$18.82    &  $-$15.86    &  $-$12.40    &   $-$11.76   \\
(mag)            &    $\pm$0.68    &   $\pm$0.65    &   $\pm$0.46    &    $\pm$0.24   \\
		 &              &            &            &            \\
 $A_{25}$        &     16.64    &    5.59    &    2.50    &     1.90   \\
(kpc)            &     $\pm$4.25    &   $\pm$1.31    &   $\pm$0.54    &    $\pm$0.22   \\
		 &              &            &            &            \\
 log[SFR]        &     $-$0.47    &   $-$1.42    &   $-$3.28    &    $-$4.56:  \\
($M_{\sun}$/yr)  &     $\pm$0.18    &   $\pm$0.31    &   $\pm$0.33    &    $\pm$0.17   \\
		 &              &            &            &            \\
 p*              &     $-$0.02    &   +0.21    &   $-$0.26    &    $-$1.29:  \\
		 &     $\pm$0.15    &   $\pm$0.14    &   $\pm$0.23    &    $\pm$0.13   \\
		 &              &            &            &            \\
 f*              &     $-$0.62    &   $-$0.48    &   +0.50    &    +0.55:  \\
		 &     $\pm$0.17    &   $\pm$0.18    &   $\pm$0.23    &    $\pm$0.18   \\
		 &              &            &            &             \\
 Log[SFR/A$^2_{25}$]    &     $-$2.77    &   $-$2.66    &   $-$3.85    &    $-$5.03:   \\
($M_{\sun}$/yr/kpc$^2$) &   $\pm$0.18      & $\pm$0.20      & $\pm$0.21      &  $\pm$0.16      \\
\hline
\end{tabular}
\end{table}

  a) The global star formation rate in the galaxies correlates well with
their luminosity, linear diameter and hydroden mass. But, being normalized
to the luminosity, the specific star formation rate, SFR/L, has appreciably
lower scatter than being normalized to the galaxy hydroden mass or
attributed to 1 $(kpc)^2$.

  b) Judging by the mean value $< p_*> = -0.02\pm0.15$, spiral galaxies in
the M~81 group would have time to generate their luminosity (baryon mass)
during the cosmological time $T_0=13.7$ billion years. About the same
conclusion may be drawn regarding the galaxies of Im, BCD types
($ < p_*> = +0.21\pm0.14$) as well as dIr galaxies ($ -0.26\pm$0.23).
However, the dSph galaxies with their $< p_*> = -1.29\pm0.13$ are capable
of reproducing only $\sim$5\% of their observed luminosity (mass).

  c) According to the mean quantities $< f_*>$, which equal $-0.62\pm$0.17
and $-0.48\pm$0.18 for S and Im,BCD galaxies, respectively, these galaxies
possess the supply of gas sufficient to maintain their observed star
formation rates for only the next (1/4 - 1/3)$T_0$ years. On the contrary,
dwarf galaxies of dIr and dSph types have the mean gas depletion time
of about 3 $T_0$.

  d) The mean specific star formation rate per 1$(kpc)^2$ is almost the same
for spiral galaxies ($-2.77\pm$0.18) as for Im, BCD types ($-2.66\pm$0.20).
However, for dIr galaxies this quantity turns out to be one order lower
($-3.85\pm$0.21), and for dSph members of the group it drops down to the
detection threshold ($-5.03\pm$0.16).

  e) All dIr and BCD galaxies with absolute magnitudes fainter than
$-17.5^m$ tend to be located on the $[p_*, f_*]$ plane along the diagonal
${f_* = -p_*}$. Because the typical error of measuring log[SFR] is rather
small ($\pm0.06)$, the observed diagonal alignement can be naturaly
explained by a stochastic burst-like variations of star formation in
this type galaxies.

  f) Curiously, in Fig.4 the quadrant $[p_*>0, f_*>0]$ is almost empty,
containing only one peculiar object, the hydrogen cloud HIJASS, with its
rather uncertain values of $p_*$ and $f_*$. It would be interesting to
search for similar objects ( dark galaxies being at a start of their
stellar evolution ?) also in other nearby groups.

  g) The data given in Table 2 show that the environment of spiral and
irregular galaxies affects slightly their star formation rate. For instance,
eight galaxies at the group outskirts with negative tidal indices have
almost the same average SFR (i.e. $p_*$) as galaxies of the same types
(S,Im,BCD, and dIr) with close massive neighbours ($TI > 0$). However, all
13 dSph galaxies with their low SFRs occur in the group core only.

\section{Conclusions}

The nearest group of galaxies around M81 appears to be a typical group
in the local universe in its population, size, velocity dispersion
and in the luminosity of the brightest member (Karachentsev, 2005).
In contrast to the Local Group, where the value of SFR for the
Milky Way remains still unknown, the values of $H\alpha$ fluxes
(or their upper limits) have been measured for all known members of the
M~81 group. Beyong the radius $R=2$ Mpc around M81, this group borders
on our Local Group and the group around the brightest galaxies
IC~342, Maffei 1, and Maffei 2. Assuming a sphere of
radius 2 Mpc around M81 to be the local ``cell of homogeneity'',
we derive the total star formation rate $\Sigma[SFR] =
(5.5\pm0.1)M_{\odot}$/year which falls within a volume of 34 Mpc$^3$.
Consequently, the density of star formation rate in this cell is
$\dot{\rho}_{SFR}$ = $ 0.165 M_{\odot}$/year Mpc$^3$.
According to Nakamura et al. (2004), Martin et al. (2005), and
Hanish et al. (2006), the average global rate of star formation
per unite volume at the present epoch (Z = 0) lies in the range
$(0.02-0.03)M_{\odot}$/year Mpc$^3$. Therefore, the M81 group demonstrates
the star formation activity 5--8 times as high as the typical neighboring
volume. Note that half of the total $H\alpha$ flux of the group falls
on one hyper-active galaxy M82. However, even after its exclusion, excess
in $\dot{\rho}_{SFR}$ for the M81 group and its surroundings is preserved.
In this sense, the statement of Miller (1996) that the star formation
process in the group M81 is highly active holds true. Possibly, the
vigorous activity of this group has two causes: the close approach of
massive galaxies M81 and M82, and the high abundance of neutral hydrogen
in the group core.

\acknowledgements{The authors are grateful to B.\ Tully, and A.\ Moiseev for useful
discussions. Support associated with HST program 10905 was provided by
NASA through a grant from the Space Telescope Science Institute,
which is operated by the Association of Universities for Research in
Astronomy, Inc., under NASA contract NAS5--26555. This work was also
supported by RFFI grant 04--02--16115.}

{}


\begin{thebibliography}{}
\bibitem{}Afanasiev V.L., Gazhur E.B., Zhelenkov S.R., Moiseev A.V., 2005,
	  Bull.SAO, 58, 90
\bibitem{} Appleton P.N., Davies R.D., Stephenson R.J., 1981, MNRAS, 195, 327
\bibitem{}Begum A., 2006,  private communication
\bibitem{}Begum A., Chengalur J.N., Karachentsev I.D., Kajsin S.S., Sharina M.E.,
	  2006, MNRAS, 365, 1220
\bibitem{}Boyce P.J., Minchin R.F., Kilborn V.A. et al., 2001, ApJ, 560, L127
\bibitem{}Feulner G., Hopp U., Botzler C.S., 2006, A\&A, 451L, 13
\bibitem{}Gallagher J.S., Hunter D.A., Tutukov A.V., 1984, ApJ, 284, 544
\bibitem{}Gil de Paz, Madore B.F., Pevunova O., 2003, ApJS, 147, 29
\bibitem{}Gutierrez C.M., Alonso M.S., Funes J.G., Ribeiro M.B., 2006,
	  AJ, 132, 596
\bibitem{}Hanish D.J., Meurer G.R., Ferguson H.C., et al., 2006, ApJ, 649, 150
\bibitem{}Hodge P.W., Kennicutt R.C., 1983, AJ, 88, 296
\bibitem{}Huchtmeier W.K., Karachentsev I.D., Karachentseva V.E., Ehle M.,
	  2000, A\&AS, 141, 469
\bibitem{}Hunter D.A., Elmegreen B.G., 2004, AJ, 128, 2170
\bibitem{}James P.A., Shane N.S., Beckman J.E., et al., 2004, A\&A, 414, 23
\bibitem{}Johnson R.A., Lawrence A., Terlevich R., Carter D., 1997, MNRAS, 287, 333
\bibitem{}Juneau S., Glazebrook K., Crampton D. et al., 2005, ApJ, 619, 135
\bibitem{}Kaisin S.S., Karachentsev I.D., 2006, Astrofizika, 49, 337
\bibitem{}Kaisin S.S., Kasparova A.V., Kniazev A.Y., Karachentsev I.D., 2006,
	  Astron. Lett. accepted
\bibitem{}Karachentsev I.D., Dolphin A.E., Tully R.B., 2006, AJ, 131, 1361
\bibitem{}Karachentsev I.D., 2005, AJ, 129, 178
\bibitem{}Karachentsev I.D., Kajsin S.S., Tsvetanov Z., Ford H., 2005, A\&A, 434, 935
\bibitem{}Karachentsev I.D., Karachentseva V.E., Huchtmeier W.K., Makarov D.I.,
	   2004, AJ, 127, 2031
\bibitem{}Karachentsev I.D., Dolphin A.E., Geisler D., et al., 2002, A\&A, 383, 125
\bibitem{}Karachentsev I.D., Karachentseva V.E., Borngen F., 1985, MNRAS,
\bibitem{}Karachentseva V.E., Karachentsev I.D., Borngen F., 1985, A\&A Supp., 60, 213
\bibitem{}Kennicutt R.C., 1998, ApJ, 498, 541
\bibitem{}Kennicutt R.C., Edgar B.K., Hodge P.W., 1989, ApJ, 337, 761
\bibitem{}Kennicutt R.C., 1989, ApJ, 344, 685
\bibitem{}Lozinskaya T.A., Moiseev A.V., Avdeev V.Yu., Egorov O.V., 2006, Astron. Lett.
       32, 361
\bibitem{}Madau P., Ferguson H.C., Dickinson M.E., et al., 1996, MNRAS, 283, 1388
\bibitem{}Makarov D.I., Karachentsev I.D., Burenkov A.N., 2003,  A\&A, 405, 951
\bibitem{}Makarova L., Grebel E., Karachentsev I. et al., in "Astrophysics and Space
       Sci., 2003, 285, 107
\bibitem{}Martin D.C., Seibert M., Buat V., et al., 2005, ApJ, 619, 59
\bibitem{}Miller B.W., 1996, AJ, 112, 991
\bibitem{}Miller B.W., Hodge P., 1994, ApJ, 427, 656
\bibitem{}Nakamura O., Fukugita M., et al., 2004, AJ, 127, 2511
\bibitem{}Schlegel, D.J., Finkbeiner, D.P., \& Davis, M., 1998, ApJ, 500, 525
\bibitem{}Spergel D.N., et al., 2003, ApJS, 148, 175
\bibitem{}Taylor E.N., Webster R.L., 2005, ApJ, 634, 1067
\bibitem{}Taylor E.N., 2005, astro-ph/0509378
\bibitem{}Tutukov A.V., 2006, Astronomy Reports, 50, 526
\bibitem{}van Zee L., 2000, AJ, 119, 2757
\bibitem{}de Vaucouleurs, G., de Vaucouleurs, A., \& Corwin, H., 1976,
 Second Reference Catalogue of Bright Galaxies, Texas, Austin
\bibitem{}Walter F., Martin C.L., Ott J., 2006, astro-ph/0608169
\bibitem{}Walter F., Skillman E.D., Brinks E., 2005, ApJ, 627, 105
\bibitem{}Walter F., Weiss A., Martin C., Scoville N., 2002, AJ, 123, 225
\bibitem{}Young J.S., Allen L., Kenney J.D., Rownd B., 1996, AJ, 112, 1903
\bibitem{}Yun M.S., Ho P.T., Lo K.Y., 1994, Nature, 372, 530
\end{thebibliography}
\end{document}